\documentclass[10pt]{iopart}

\usepackage{iopams}
\usepackage{graphicx, bm, color}

\newcommand{\kappat}{\tilde{\kappa}}

\newcommand{\rv}{\mathbf{r}}
\newcommand{\Rv}{\mathbf{R}}
\newcommand{\ev}{\mathbf{e}}

\newcommand{\uv}{\mathbf{u}}
\newcommand{\xv}{\mathbf{x}}

\begin{document}

\title[]{Elasticity of randomly diluted honeycomb and diamond lattices with bending forces}

\author{Danilo B. Liarte}
%\ead{dl778@cornell.edu}
\address{Laboratory of Atomic and Solid State Physics, Cornell University, Ithaca, NY, USA}
\address{Institute of Physics, University of S\~ao Paulo, S\~ao Paulo, SP, Brazil}
\author{O. Stenull}
%\ead{olaf.stenull@comcast.net}
\address{Department of Physics and Astronomy, University of Pennsylvania, Philadelphia, PA, USA}
\author{Xiaoming Mao}
%\ead{maox@umich.edu}
\address{Department of Physics, University of Michigan, Ann Arbor, MI, USA}
\author{T. C. Lubensky}
%\ead{tom@physics.upenn.edu}
\address{Department of Physics and Astronomy, University of Pennsylvania, Philadelphia, PA, USA}

\vspace{10pt}
\begin{indented}
\item[]January 2016
\end{indented}

\begin{abstract}
We use numerical simulations and an effective-medium theory to
study the rigidity percolation transition of the honeycomb and
diamond lattices when weak bond-bending forces are included.
We use a rotationally invariant bond-bending potential,
which, in contrast to the Keating potential, does not involve
any stretching. As a result, the bulk modulus does not depend
on the bending stiffness $\kappa$. We obtain scaling functions
for the behavior of some elastic moduli in the limits of small
$\Delta \mathcal{P} =  1 - \mathcal{P}$, and small $\delta
\mathcal{P} = \mathcal{P} - \mathcal{P}_c$, where $\mathcal{P}$
is an occupation probability of each bond, and
$\mathcal{P}_c$ is the critical probability at which rigidity
percolation occurs. We find good quantitative agreement between
effective-medium theory and simulations for both lattices for
$\mathcal{P}$ close to one.
\end{abstract}

\submitto{\JPCM}
%\maketitle
\ioptwocol

\section{Introduction}
Concepts associated with the rigidity percolation transition of
random elastic networks \cite{thorpe83,feng84a} have been
applied in many branches of science, such as amorphous solids
\cite{alexander98, wyart05}, granular materials
\cite{edwards99, tkachenko99}, mineralogy \cite{hammonds96},
networks of semi-flexible polymers
\cite{BroederszMac2014,Broedersz2011} and the mechanics of
living cells \cite{das12,Lenz2014,RoncerayLenz2015} . The
archetype of this transition \cite{feng85, he85,
mao11,Broedersz2011,mao13a, mao13b} occurs in periodic lattices
in which bonds consisting of central-force springs are
populated with probability $\mathcal{P}$.  For $\mathcal{P}$
below a threshold $\mathcal{P}_c$, the lattice loses rigidity,
and some or all of its elastic moduli vanish because floppy
regions prevent rigid ones from percolating.

\begin{figure}[!ht]
(a) \par\smallskip
\centering
\includegraphics[width=0.6\linewidth]{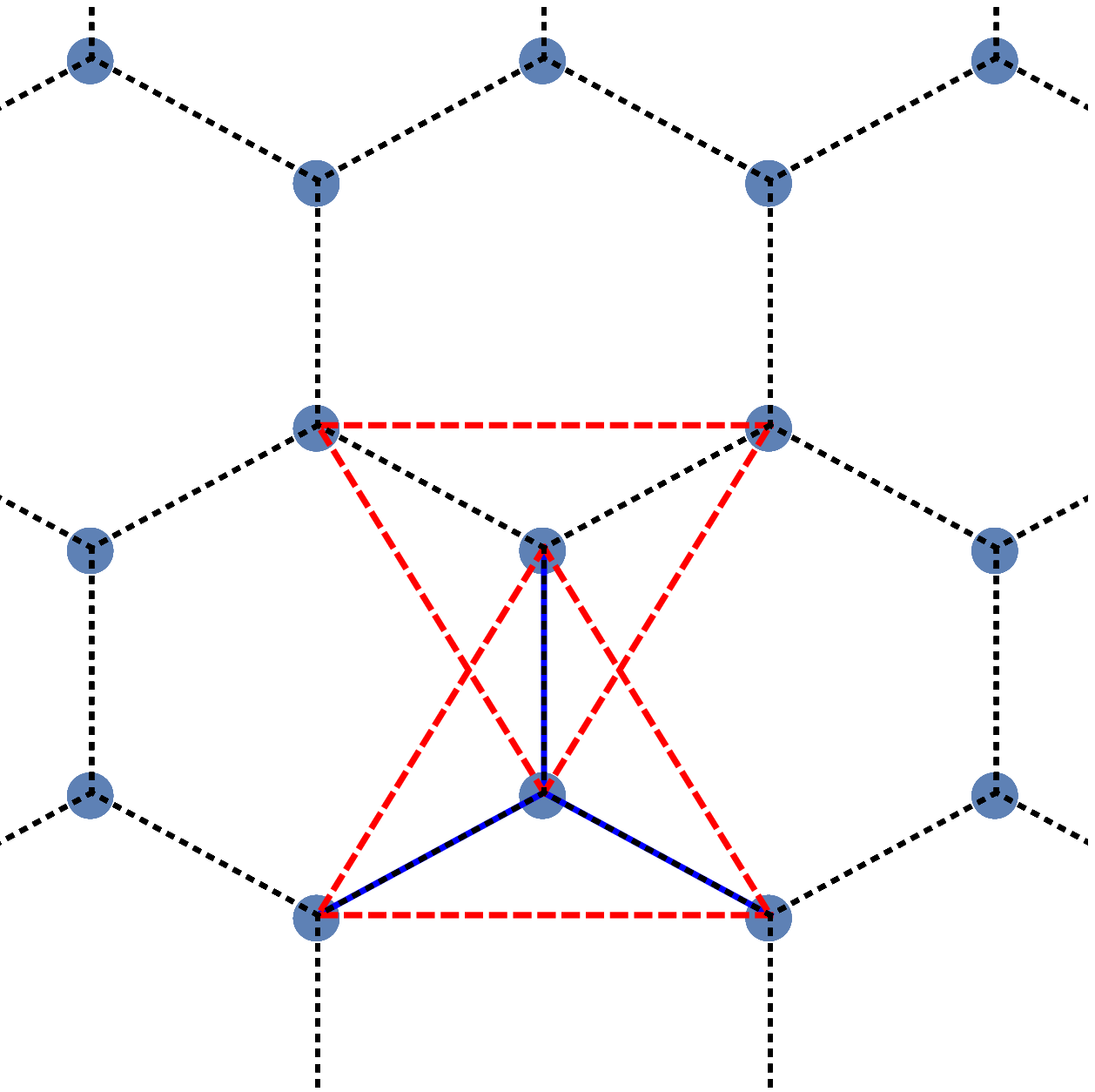}
\\
(b) \par\smallskip
\centering
\includegraphics[width=0.7\linewidth]{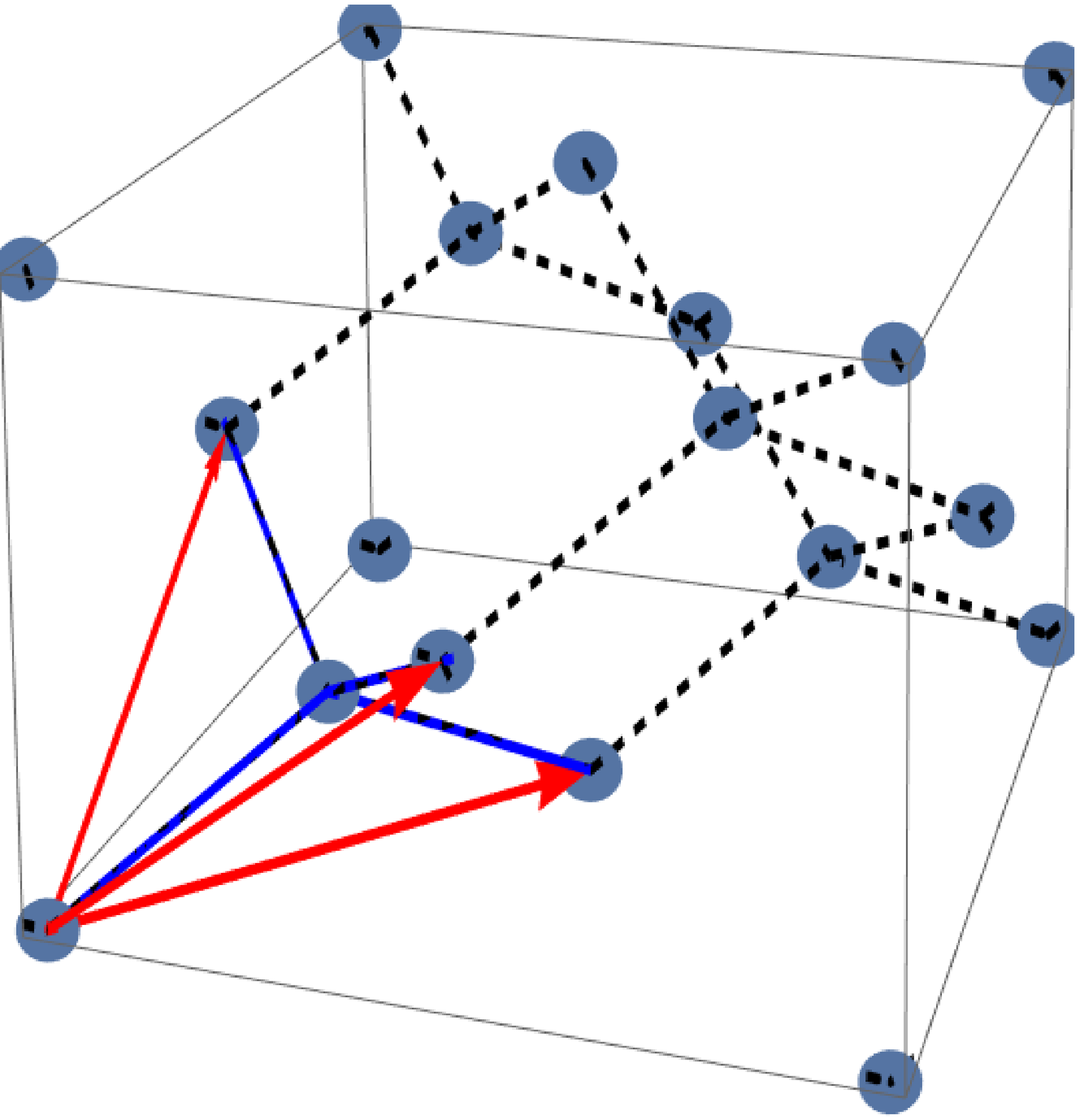}
\caption{a) Crystal structure of the honeycomb net. b) Conventional cubic cell of the diamond lattice.%
\label{honeyDiam}}
\end{figure}

The homogeneous honeycomb and diamond lattices
(Fig.~\ref{honeyDiam}) with only nearest-neighbor bonds are
strongly \emph{under-coordinated}. The average coordination number $z$ of both the honeycomb and
diamond lattices is less than the Maxwell limit
\cite{maxwell65, souslov09} $z_c = 2 d$ for central forces,
where $d$ is the spatial dimension, below which lattices under
periodic boundary conditions develop zero-frequency ``floppy"
modes. The honeycomb lattice with $2$ sites per unit cell and
$z=3$ has a deficiency of one bond, and the diamond lattice
with $2$ sites per unit cell and $z=4$ has a deficiency of two
bonds per unit cell relative to the Maxwell limit. As a result, they have an
extensive number of zero modes and no resistance to shear
distortions; but curiously because of the special geometry of
their lattices, they both have a non-zero bulk modulus $B$.
Thus extra forces, such as next-nearest-neighbor central forces
or bending forces favoring a particular angle between pairs of
bonds sharing common endpoints, are required for mechanical
stability. Here, using both effective medium theory (EMT) and
numerical simulations, we study the properties of randomly
bond-diluted central-force honeycomb and diamond lattices with
added bending forces characterized by a local bending stiffness
$\kappa$, focussing in particular on behavior near zero
dilution ($\mathcal{P}\approx 1$ ) and near the rigidity
threshold at $\mathcal{P}=\mathcal{P}_c$. We find that the bulk
modulus of both lattices can be expressed near $\mathcal{P}=1$
as a scaling function of $\kappa/(1-\mathcal{P})^n$ with $n=1$,
much like the shear modulus in the diluted kagome lattice with
bending forces \cite{mao13a} where $n=2$ rather than $1$. The
shear moduli of the honeycomb and diamond lattices, on the
other hand exhibit, no simple scaling form and approach zero
even at $\mathcal{P}=1$ as $\kappa \rightarrow 0$ as required.
We find that the bulk moduli, much like the shear moduli in the
Mikado model \cite{Head2003,Head2003a,Wilhelm2003} and the
diluted triangular \cite{Broedersz2011} and kagome lattices
with bending \cite{mao13a},  exhibit crossover from stretching
dominated affine response to bending dominated nonaffine
response with decreasing $\mathcal{P}$, reaching a maximum
nonaffinity at the rigidity threshold.  The shear modulus, which
vanishes with $\kappa$, on the other hand, always exhibits
bending and thus non-affine response.

Bending
forces effectively couple next-nearest neighbor sites,
thereby increasing the effective $z$ to values above the
central-force critical value, $z_c$, providing elastic
stability, and eliminating all but the trivial zero modes of
rigid translation and rotation. In contrast to Keating
potentials \cite{keating66}, bending forces do not depend on
the length of bonds and, therefore, do not contribute to the
bulk modulus at $\mathcal{P} = 1$. The Keating potential is
rotational invariant. Bending forces are also, as we show in
the Appendix.

\section{Model}

The  two-dimensional honeycomb lattice (Fig.~\ref{honeyDiam}a)
is a triangular Bravais lattice with a two-point basis
\cite{ashcroft76}.  It is defined by a set of primitive
vectors, e.g. $\bm{a}_1 = \sqrt{3} (1,0)$, and $\bm{a}_2 =
(\sqrt{3}/2) (1, \sqrt{3})$, along with the positions of the
atoms within each primitive cell, $\bm{c}_1 = (0,0)$, and
$\bm{c}_2 = (0,1)$. The length of each bond in the lattice
is $\ell_0=1$. The three-dimensional diamond lattice can be
represented as a face-centered cubic lattice with a two-point
basis (Fig.~\ref{honeyDiam}b). It can be defined by the
primitive vectors $\bm{a}_1 = (1/2) (0,1,1)$, $\bm{a}_2 = (1/2)
(1,0,1)$, $\bm{a}_3 = (1/2) (1,1,0)$, along with the two-point
basis vectors $\bm{c}_1 = (0,0,0)$, and $\bm{c}_2 = (1/4)
(1,1,1)$. Bonds connecting nearest-neighbor sites are of length
to $\ell_0=\sqrt{3}/4$.

We consider the following lattice energy of interaction:
\begin{eqnarray}
E = E_{\mathrm{stretch}} + E_{\mathrm{bend}}.
\label{eq:E-sb}
\end{eqnarray}
The first term is a sum of central-force interactions between
nearest-neighbor pairs of sites $i$ and $j$, which in the
harmonic limit are given by:
\begin{eqnarray}
E_{\mathrm{stretch}}^{ij}=\frac{1}{2} k \left[\left(\bm{u}_j-\bm{u}_i\right)
\cdot \bm{\hat{r}}_{ij}\right]^2,
\label{eq:Estr}
\end{eqnarray}
where $\bm{\hat{r}}_{ij}$ is the unit vector connecting sites
$i$ and $j$ in the undeformed lattice, and $\bm{u}_i$ is a
displacement vector. Each unit cell in the honeycomb lattice
has three independent bonds (e.g., the blue lines of Fig.
\ref{honeyDiam}a), and in the diamond lattice, four independent
bonds. The second term in Eq.~(\ref{eq:E-sb}) is a sum of
bending energy interactions associated with two bonds,
terminating on a common site $l$ and connecting
nearest-neighbor pairs of sites $(l,m)$, and $(l,k)$:
\begin{eqnarray}
E_{\mathrm{bend}}^{klm}
%\equiv \frac{ \kappat \, k \, \ell_0^2 }{2} \left(\sin \beta_0 \Delta \beta_{klm} \right)^2,
\equiv \frac{ \kappa }{2} \left(\sin \beta_0 \Delta \beta_{klm} \right)^2,
\label{ebend}
\end{eqnarray}
where
% $\kappat = \kappa/(k \ell_0^2)$ is a unitless measure of the bending stiffness,
$\beta_0$ is the equilibrium angle between
bonds, and $\Delta \beta_{klm}$ represents the difference between the angle between the bonds $(l,m)$ and $(l,k)$ and $\beta_0$. There are six such terms per primitive cell for the honeycomb
lattice, and twelve for the diamond lattice. The bending
interactions effectively couple next-nearest-neighbors sites,
as illustrated by the red dashed lines of
Fig.~\ref{honeyDiam}a. The factor $\sin \beta_0$ is a matter of
convenience. It is equal to $\sqrt{3}/2$ for the honeycomb
lattice, and $2\sqrt{2}/3$ for the diamond lattice. Equation
(\ref{ebend}) may be written as a combination of displacement
vectors, up to second order in $\bm{u}$, as
\begin{eqnarray}
E_{\mathrm{bend}}^{klm}
& = & \frac{\kappa}{2 \ell_0^2} \Big\{ \bm{u}_{lm} \cdot \left[\bm{\hat{r}}_{lk}
%& = & \frac{1}{2} \kappat k \Big\{ \bm{u}_{lm} \cdot \left[\bm{\hat{r}}_{lk}
-(\bm{\hat{r}}_{lk}\cdot \bm{\hat{r}}_{lm})\bm{\hat{r}}_{lm}\right] \nonumber \\
&& \quad +
\bm{u}_{lk} \cdot \left[\bm{\hat{r}}_{lm}-(\bm{\hat{r}}_{lk}\cdot \bm{\hat{r}}_{lm})\bm{\hat{r}}_{lk}\right] \Big\}^2,
\label{ebend-b}
\end{eqnarray}
where $ \bm{u}_{lk} = \bm{u}_k - \bm{u}_l$.

\begin{figure}[th]
\vspace{0.5cm} \begin{minipage}[b]{0.38\linewidth}
(a) \par\smallskip
\centering
\includegraphics[width=\linewidth]{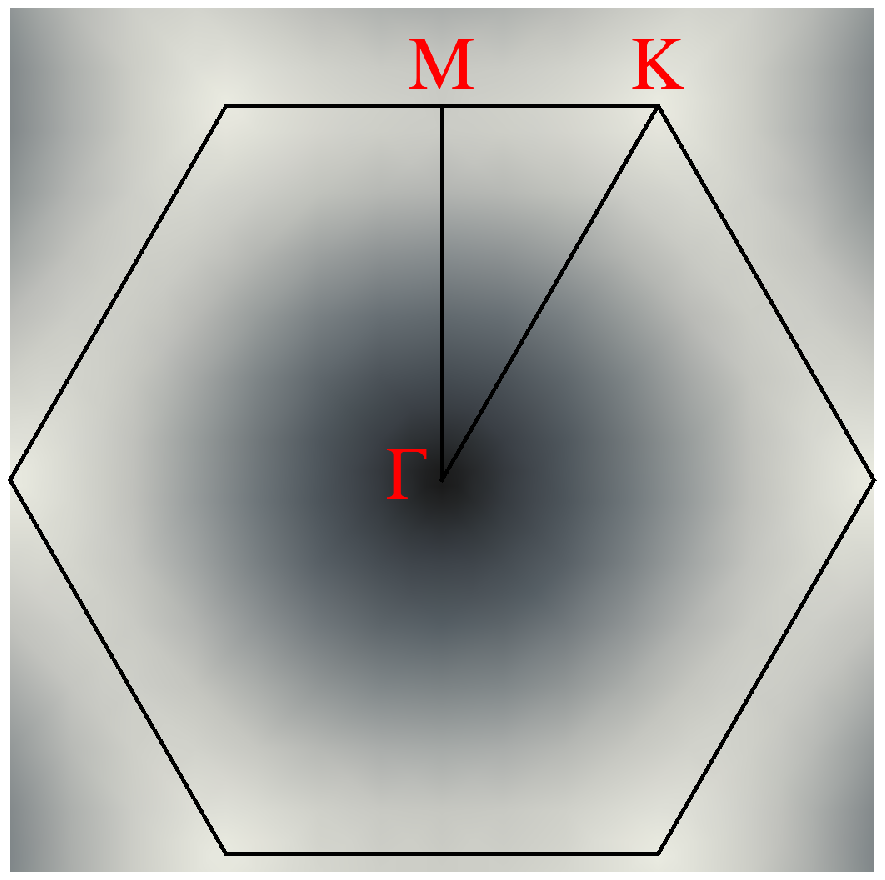}
\end{minipage}
\hspace{0.1cm} \begin{minipage}[b]{0.58\linewidth}
(b) \par\smallskip
\centering
\includegraphics[width=\linewidth]{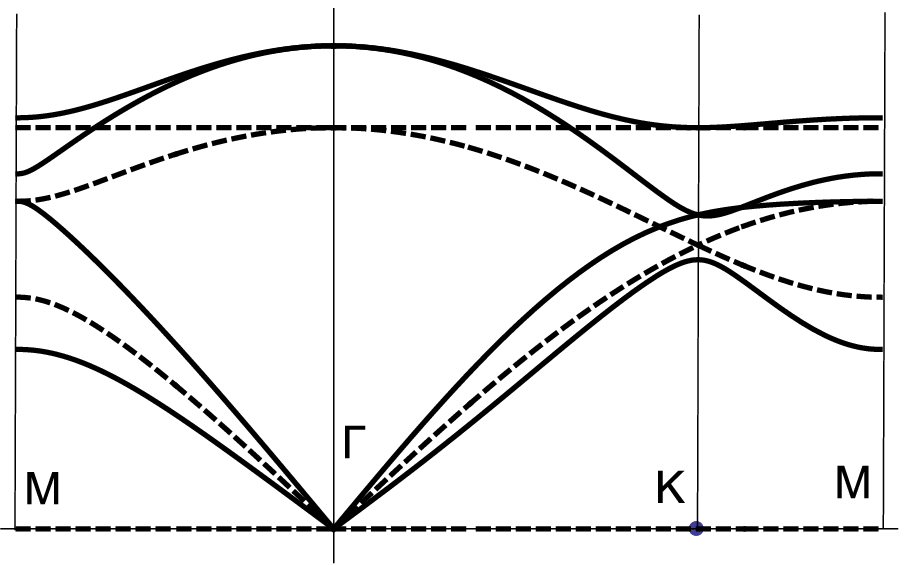}
\end{minipage}
\caption{a) Density plot in the $q_x\times q_y$ plane of one of the honeycomb's
acoustic modes for $k=1$, and $\kappa = 0.01$. b) Dispersion curves
along some symmetry lines for $k=1$, $\kappa = 0$ (dashed curves),
and $\kappa = 0.1$ (solid curves).}%
\label{dispersion1Honeycomb}%
\end{figure}

\begin{figure}[th]
\vspace{0.5cm}
\par
\begin{center}
\includegraphics[width=0.8\linewidth]{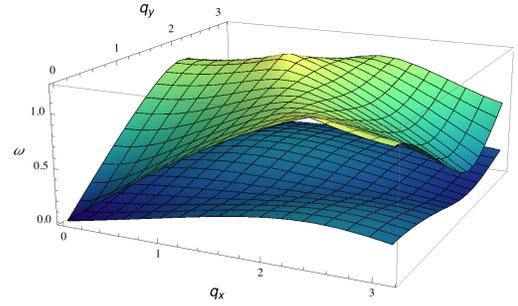}
\end{center}
\caption{Three-dimensional plot of the acoustic branch
frequencies of the honeycomb lattice for $k=1$ and $\kappa=0.01$.}%
\label{dispersion2Honeycomb}
\end{figure}

\begin{figure}[th]
\begin{minipage}[b]{0.65\linewidth}
\includegraphics[width=\linewidth]{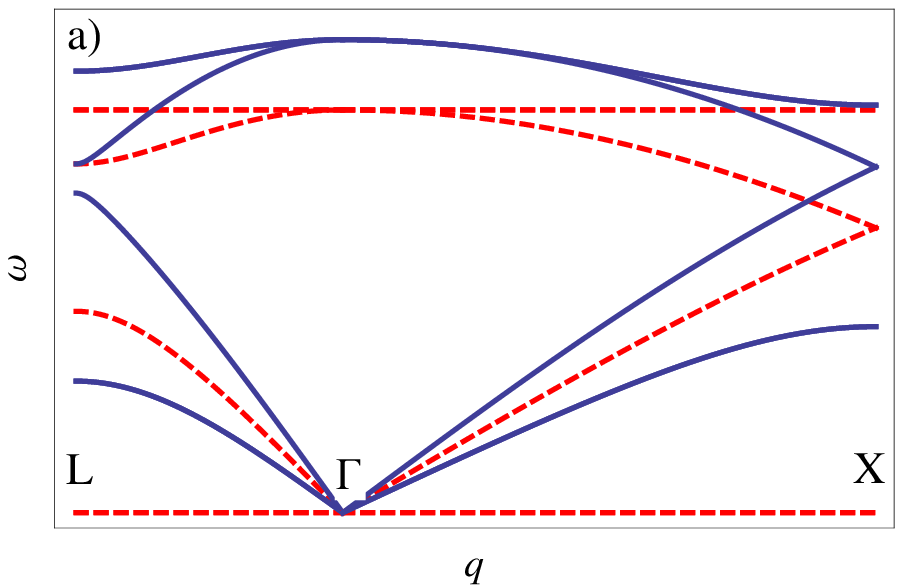}
\end{minipage}
\hspace{0.5cm}
\begin{minipage}[b]{0.25\linewidth}
\includegraphics[width=\linewidth]{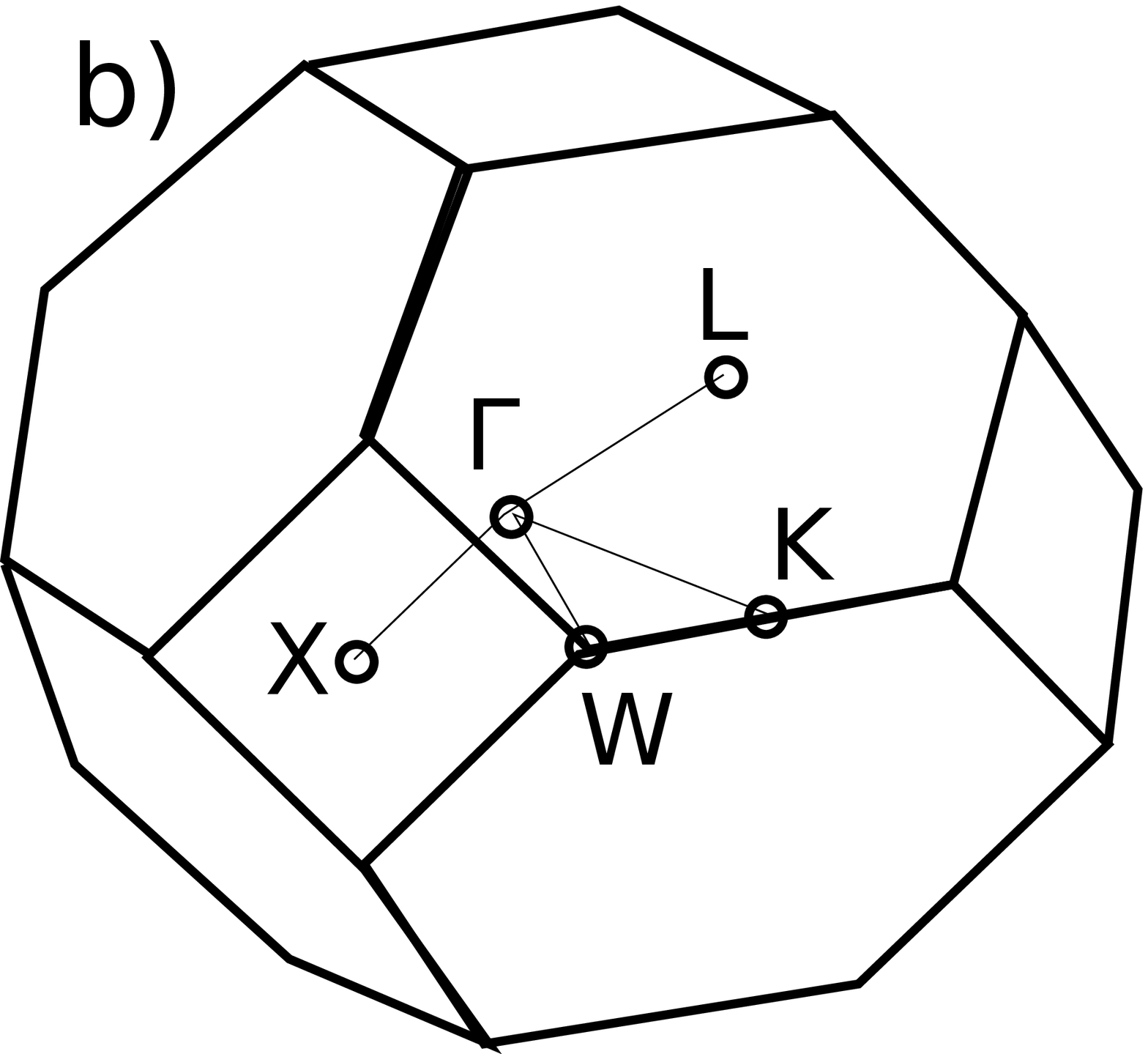}
\vspace{0.5cm}
\end{minipage}
\caption{a) Dispersion curves of the diamond lattice along symmetry lines
$L-\Gamma$ and $\Gamma-X$ The dashed lines represent $\kappa = 0$ and the
full lines $\kappa = 0.01$. b) Sketch of the first Brillouin zone of the
diamond lattice and high symmetry points $\Gamma$, $L$, $X$, $K$, and $W$.}%
\label{dispersionDiamond}%
\end{figure}

The honeycomb lattice has four phonon branches, corresponding
to the four degrees of freedom within each unit cell. In the
limit $\kappa \rightarrow 0$, one of the two acoustic branches
is floppy with zero frequency for all wavenumbers
$\bm{q}=(q_x,q_y)$ in the Brillouin Zone. Figure
\ref{dispersion1Honeycomb}a shows a density plot of one of the
acoustic branch frequencies for homogeneous $k=1$, and for
$\kappa=\kappat=0.01$. Hereafter $k=1$ is assumed unless
otherwise noted. Figure \ref{dispersion1Honeycomb}b displays
dispersion curves along symmetry lines $\Gamma M$, $\Gamma K$,
and $KM$, $\kappa = 0$ (dashed curves), and $\kappa = 0.1$
(solid curves). Figure \ref{dispersion2Honeycomb} show a $3d$
plot of the two acoustic branches as a function of $q_x$ and
$q_y$, for $\kappa=0.01$. The diamond lattice has six phonon
branches, of which two are floppy when $\kappa \rightarrow 0$.
Figure \ref{dispersionDiamond}a shows diamond-lattice
dispersion curves for $\kappa=0.01$, along symmetry lines
$\Gamma X$ and $\Gamma L$. A sketch of the first Brillouin zone
of the diamond lattice with five high symmetry points is
displayed in Figure \ref{dispersionDiamond}b. Notice that two
largest and the two smallest eigenvalues are degenerate in both
lines. These degeneracies can be broken along other less
symmetrical lines.

\section{Effective-medium theory and simulations}

We use an adaptation of the EMT developed in
Ref.~\cite{das07,das12} in which the spring constants of
individual bonds and the bending constants of individual bond
pairs are treated as independent random variables~\footnote{This version of CPA is widely used and accepted in treatments of rigidity percolation. We divide the interactions up into equivalency classes and average, so our effective Hamiltonian is writable as a sum of homomorphic parts. This version of CPA is equivalent to what Yonezawa and Ogadaki call the HCPA~\cite{yonezawa78}, which yields an analytical physical solution.}.  The
probability that a given bond is occupied is $\mathcal{P}$, and
the probability distribution for the spring constant for any
bond is
\begin{equation}
P_{\mathrm{spr}} (k')= \mathcal{P} \delta(k'-k) + (1-\mathcal{P}) \delta(k') .
\end{equation}
Both bonds of a bond pair must be occupied in order for there
to be a bending energy associated with the pair. Following the
approximation of Refs.~\cite{das07,das12}, we set the
probability that a given bond pair exists equal to $\mathcal{P}^2$.
The probability distribution for the bending constant of an
individual bond pair is
\begin{equation}
P_{\mathrm{bend}}= \mathcal{P}^2 \delta(\kappa'-\kappa)+(1-\mathcal{P}^2)\delta(\kappa') .
\end{equation}
The joint probability for both $k'$ and $\kappa'$ is then
$P(k',\kappa') = P_{\mathrm{spr}}(k') P_{\mathrm{bend}}(\kappa')$.

In this EMT theory, each occupied bond and each occupied bond
pair constitutes a constraint.  Thus if each site has $z$
neighbors, there are $(z\mathcal{P}/2) N$ bond constraints and
$(z(z-1)\mathcal{P}^2/2) N$ bond-pair constraints in a diluted
lattices of $N$ sites.  Since each site has $d$ translational
degrees of freedom, the Maxwell count of the number of zero
modes per site is
\begin{equation}
f = d - \frac{z\mathcal{P}}{2} - \frac{z (z-1)}{2} \mathcal{P}^2,
\label{floppy_counting} .
\end{equation}
The EMT rigidity threshold is obtained by setting $f=0$ to
produce
\begin{equation}
\mathcal{P}_c = \frac{1}{2(z-1)} \left[\sqrt{1 + \frac{8d(z-1)}{z}}  - 1\right] .
\label{Pcrit}
\end{equation}
Thus, in the honeycomb lattice $\mathcal{P}_c \approx 0.60$ and
the average coordination number at threshold, $z_c=z
\mathcal{P}$, is $1.8$; in the diamond lattice, $\mathcal{P}_c
\approx 0.56$ and $z_c \approx 2.24$.

The EMT threshold should be compared with two other estimates
of $z_c$ based on the Maxwell zero-mode count. Phillips
\cite{phillips79,phillips81}, who considered general off
lattice networks, also treated bond pairs as independent and
obtained an equation identical to Eq.~(\ref{floppy_counting})
but without the $\mathcal{P}$ factors and with $z\rightarrow r$
interpreted as the average coordination number, $r=\mathcal{P}
z$. His estimate leads to $r_c = \sqrt{2d}$ or $r_c = 2$
($\mathcal{P}_c = 0.67$) and $r_c = 2.45$ ($\mathcal{P}=0.61$)
for the honeycomb and diamond lattices, respectively. Thorpe
later showed \cite{thorpe83} that treating each bond pair
independently over-counts the number of bending constraints. In
his analysis, a single pair of bonds, sharing a common site,
contributes one constraint due to bending forces. Each
additional bond sharing that site is constrained to have a
particular orientation, and thus adds a total of $d-1$
constraints, which is the number of angles needed to specify a
unit vector in $d$-dimensions. Therefore, the total number of
bending-force constraints (for $r\geq 3$) reads,
\begin{eqnarray}
&& n_{\mathrm{B}}^{\mathrm{Th}}(r+1) = d-1 + n_{\mathrm{B}}^{\mathrm{Th}}(r) \Rightarrow \\
&& n_{\mathrm{B}}^{\mathrm{Th}}(r) = (d-1) r - (2d-3),
\label{eq:bConstraintsTh}
\end{eqnarray}
with
\begin{eqnarray}
r= \left( \sum_{i=2}^z i \, n_i \right) \left/ \left( \sum_{i=2}^z n_i \right. \right),
\label{eq:rThorpe}
\end{eqnarray}
where $n_i$ is the number of $i$-coordinated sites. Note that
in applying Eq.~(\ref{eq:bConstraintsTh}), one has to ensure
that there are neither isolated nor $1$-coordinated sites.
Thorpe's counting, including $r/2$ stretching
constraints, gives $r_c= 6(d-1)/(2d -1)=2$ and $12/5=2.4$ for
the honeycomb and diamond lattices respectively.

In effective-medium theory, the diluted lattice is modeled by a
homogeneous lattice with stiffness constants $k_m$ and
$\kappa_m$ satisfying a set of self-consistent equations that
depend on a probability distribution $P(k^\prime,
\kappa^\prime)$.  We use an adaptation of the effective-medium
theories proposed in Ref.~\cite{feng85} to derive the set of
effective-medium theory equations for $k_m$ and $\kappa_m$
\cite{das07,das12,mao13a}:
\begin{eqnarray}
\frac{k_m}{k} & = &\frac{\mathcal{P}-a^*}{1-a^*},
\label{eq:k_m} \\
%\end{eqnarray}
%\begin{eqnarray}
\frac{\kappa_m}{\kappa}&=&\frac{\mathcal{P}^2-b^*}{1-b^*},
\label{eq:kappa_m}
\end{eqnarray}
where
\begin{eqnarray}
a^*=\frac{1}{z} \int_{1BZ} \frac{d \bm{q}}{\tilde{v}_0} \mathrm{Tr} \left(\bm{D}_{s,\bm{q}} \cdot \bm{D}_{\bm{q}}^{-1} \right),
\label{aStar}
\end{eqnarray}
\begin{eqnarray}
b^*=\frac{1}{(z-1) z} \int_{1BZ} \frac{d \bm{q}}{\tilde{v}_0} \mathrm{Tr} \left(\bm{D}_{b,\bm{q}} \cdot \bm{D}_{\bm{q}}^{-1} \right),
\label{bStar}
\end{eqnarray}
where $\tilde{v}_0$ is the volume of the first Brillouin zone.
In Eqs.~(\ref{aStar}) and (\ref{bStar}), $\bm{D}_{\bm{q}}$ is
the translational invariant dynamical matrix in Fourier space:
\begin{eqnarray}
\bm{D}_{\bm{q}, \bm{q}^\prime} = N \delta_{\bm{q}, \bm{q}^\prime} \bm{D}_{\bm{q}},
\end{eqnarray}
\begin{eqnarray}
\bm{D}_{\bm{q}} = \bm{D}_{s, \bm{q}} + \bm{D}_{b, \bm{q}},
\label{dynamicalMatrix}
\end{eqnarray}
where $\bm{D}_{s, \bm{q}}$, and $\bm{D}_{b, \bm{q}}$ are
contributions from the stretching and bending interactions
respectively. They may be written as
\begin{eqnarray}
\bm{D}_{s, \bm{q}} = k_m \sum_{n=1}^z \bm{B}_{n,\bm{q}}^s\bm{B}_{n,-\bm{q}}^s,
\label{dynamicalMatrixS}
\end{eqnarray}
where,
\begin{eqnarray}
\bm{B}_{n,\bm{q}}^s = \left\{e^{-i\bm{q}\cdot \bm{f}_n} \bm{e}_n, - \bm{e}_n \right\},
\label{bVectorsS}
\end{eqnarray}
and,
\begin{eqnarray}
\bm{D}_{b, \bm{q}} & = &\frac{\kappa_m}{{\ell_0}^2} \sum_{1\leq m < n \leq z} \left( \bm{B}_{mn,\bm{q}}^{b\, (i)} \bm{B}_{mn,-\bm{q}}^{b\, (i)}
\right. \nonumber \\ && \quad \left.
 +
\bm{B}_{mn,\bm{q}}^{b\, (ii)}\bm{B}_{mn,-\bm{q}}^{b\, (ii)}
 \right), 
% \nonumber \\ & &
\label{dynamicalMatrixB}
\end{eqnarray}
where $\ell_0$ is the lattice spacing, and
\begin{eqnarray}
\bm{B}_{mn,\bm{q}}^{b\, (i)} &=& \left\{e^{-i \bm{q} \cdot \bm{f}_m} \bm{e}_{nm}^{\perp}
+e^{-i \bm{q} \cdot \bm{f}_n} \bm{e}_{mn}^{\perp},
\right. \nonumber \\ && \quad \left.
-(\bm{e}_{mn}^{\perp}+\bm{e}_{nm}^{\perp}) \right\}, 
%\nonumber \\
%& & \\
\label{bVectorsB1} \\
%\end{eqnarray}
%\begin{eqnarray}
\bm{B}_{mn,\bm{q}}^{b\, (ii)} &= &\left\{
-(\bm{e}_{mn}^{\perp}+\bm{e}_{nm}^{\perp}), e^{i \bm{q} \cdot
\bm{f}_m} \bm{e}_{nm}^{\perp} 
\right. \nonumber \\ && \quad \left.
+e^{i \bm{q} \cdot \bm{f}_n}
\bm{e}_{mn}^{\perp} \right\}.
\label{bVectorsB2}
\end{eqnarray}
The vectors $\bm{e}_n$ and $\bm{f}_m$ connect nearest-neighbor
sites and cells, respectively, \footnote{We have chosen
$\bm{e}_1 = -\bm{c}_2$, $\bm{e}_2 = \bm{a}_2 -\bm{c}_2$,
$\bm{e}_3 = \bm{a}_2-\bm{a}_1-\bm{c}_2$, $\bm{f}_1 = \bm{c}_1$,
$\bm{f}_2 = \bm{a}_2$, and $\bm{f}_3 = \bm{a}_2 - \bm{a}_1$ for
the honeycomb lattice, and $\bm{e}_1 = \bm{a}_1-\bm{c}_2$,
$\bm{e}_2 = \bm{a}_2-\bm{c}_2$, $\bm{e}_3 = \bm{a}_3-\bm{c}_2$,
$\bm{e}_4 = -\bm{c}_2$, $\bm{f}_1 = \bm{a}_1$, $\bm{f}_2 =
\bm{a}_2$, $\bm{f}_3 = \bm{a}_3$, $\bm{f}_4 = \bm{c}_1$ for the
diamond lattice.}, and
\begin{eqnarray}
\bm{e}_{mn}^{\perp}= \bm{e}_m - (\bm{e}_m \cdot \bm{e}_n) \bm{e}_n.
\end{eqnarray}
These definitions imply that $a^*$ and $b^*$ are functions of
the dimensionless ratio:
\begin{eqnarray}
\tilde{\kappa}_m \equiv \frac{\kappa_m}{k_m \, \ell_0^2},
\label{eq:definition_bm}
\end{eqnarray}
rather than of $\kappa_m$ and
$k_m$ separately. Equations (\ref{aStar}), (\ref{bStar}), and
(\ref{dynamicalMatrix}) lead to the following important
relation between $a^*$ and $b^*$:
\begin{eqnarray}
a^*+ (z-1) \, b^*=(2d)/z.
\label{abRelation}
\end{eqnarray}
At the rigidity threshold, $k_m=0$ and $\kappa_m = 0$.
Equations (\ref{eq:k_m}) and (\ref{eq:kappa_m}) then require
$a^* = \mathcal{P}$ and $b^*=\mathcal{P}^2$, and
Eq.~(\ref{abRelation}) reduces to Eq,~(\ref{floppy_counting})
at $f=0$ and yields Eq.~(\ref{Pcrit}) for $\mathcal{P}_c$.

\subsection{Simulations}
In the numerical portion of our work, we generate diluted
honeycomb and diamond lattices on a computer. The systems sizes
that we simulate range up to $100^2$ unit cells for the
honeycomb lattice and $20^3$ unit cells for the diamond
lattice. In all our simulations, periodic boundary conditions
are applied. To facilitate the computations, we split up the
elastic displacement $\bm{u}_i$ into an affine and a non-affine
part,
\begin{equation}
\bm{u}_i = \eta \bm{x}_i + \delta \bm{u}_i \, ,
\end{equation}
where $\bm{x}_i$ is the equilibrium position of site $i$ in the
absence of any applied deformation, $\eta$ is the deformation
gradient tensor, and $\delta \bm{u}_i$ is the non-affine
displacement. We fix the non-affine displacement of an
arbitrarily chosen lattice site to be zero so that spurious
zero modes associated with rigid translations of the lattice
are suppressed. We apply shear and bulk deformations by
choosing $\eta$ accordingly. For example, to apply shear to the
honeycomb lattice, we chose the 2 diagonal components of $\eta$
to be zero and its 2 off-diagonal elements to be equal to
$\gamma$, where $\gamma$ is the magnitude of the deformation
which we set to $\gamma = 0.01$. Then we relax the $\delta
\bm{u}_i$ using a standard conjugate gradient algorithm that
provides us with the equilibrium non-affine displacements
$\delta \bm{u}_i^{\mathrm{na}}$ in the presence of applied
deformation. Feeding these back into the elastic model energy
density (1), we obtain the shear and bulk moduli as function of
$\mathcal{P}$ and $\kappa$. In addition to the elastic moduli,
we also compute the so-called non-affinity parameter $\Gamma$
which measures the degree of non-affinity in the system under
the applied deformation,
\begin{equation}
\Gamma = \frac{1}{N\, \gamma^2}\sum_i \left( \delta \bm{u}_i^{\mathrm{na}} \right)^2  \, ,
\end{equation}
where $N$ is the total number of sites. Our numerical results
will be displayed and discussed together with our EMT results
as we move along.

\section{Results}
%\subsection{Honeycomb lattice}

\begin{figure*}[!th]
\vspace{0.5cm} \begin{minipage}[b]{0.48\linewidth}
\includegraphics[width=\linewidth]{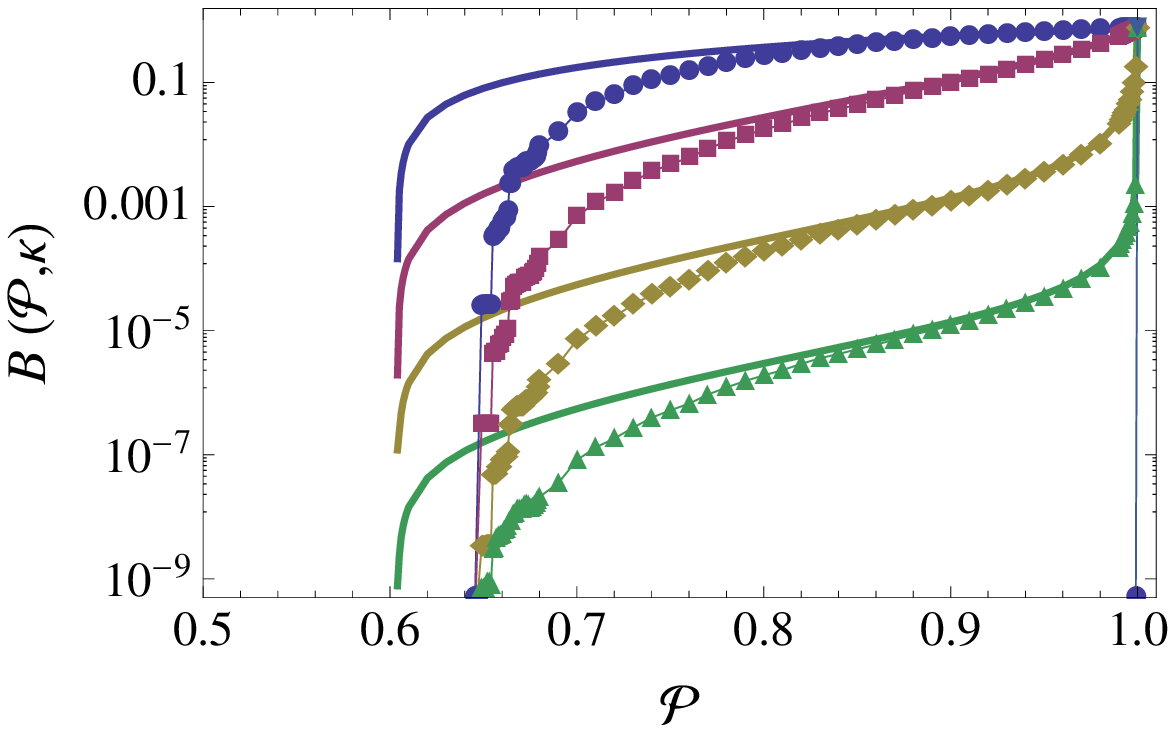}
\end{minipage}
\hspace{0.5cm} \begin{minipage}[b]{0.48\linewidth}
\includegraphics[width=\linewidth]{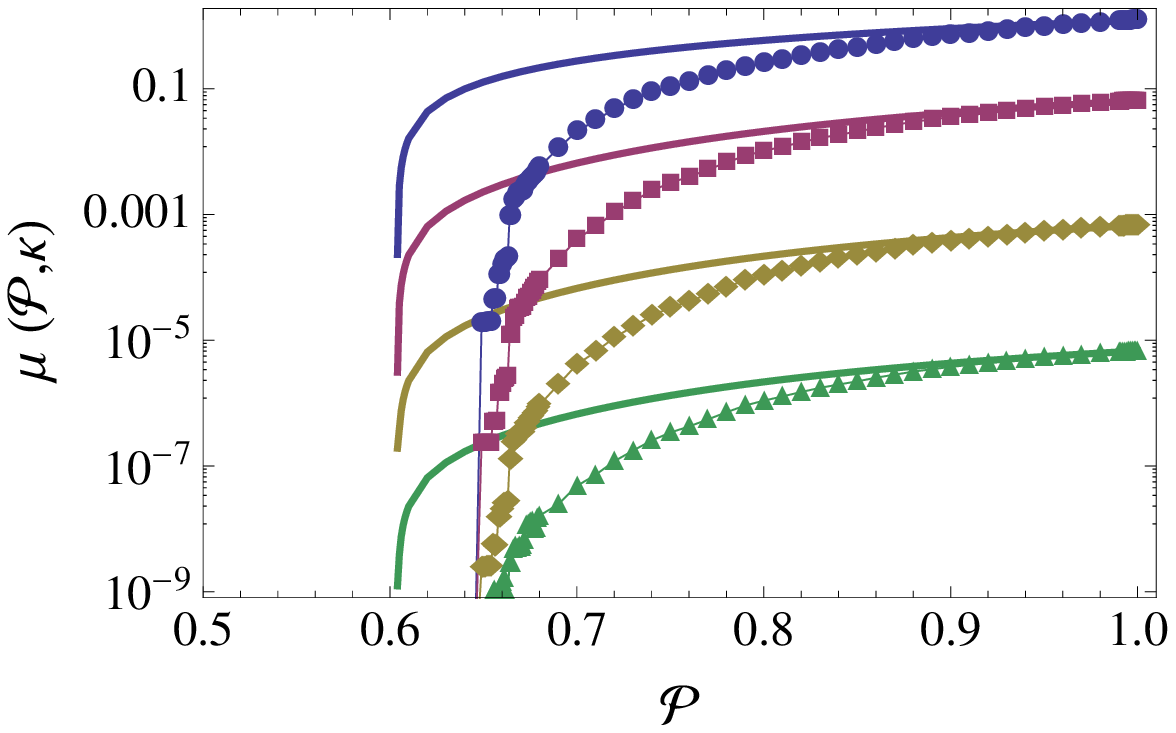}
\end{minipage}
\caption{Bulk (on the left) and shear (on the right) moduli of the diluted
honeycomb lattice as a function of $\mathcal{P}$, for $k=1$, and
$\kappa = 1$ (blue circles), $10^{-2}$ (red squares), $10^{-4}$ (yellow diamonds), and $10^{-6}$ (green triangles), from both simulations (symbols), and effective-medium theory (solid lines).}%
\label{bmup}%
\end{figure*}

\begin{figure*}[th]
\vspace{0.5cm} \begin{minipage}[b]{0.48\linewidth}
\includegraphics[width=\linewidth]{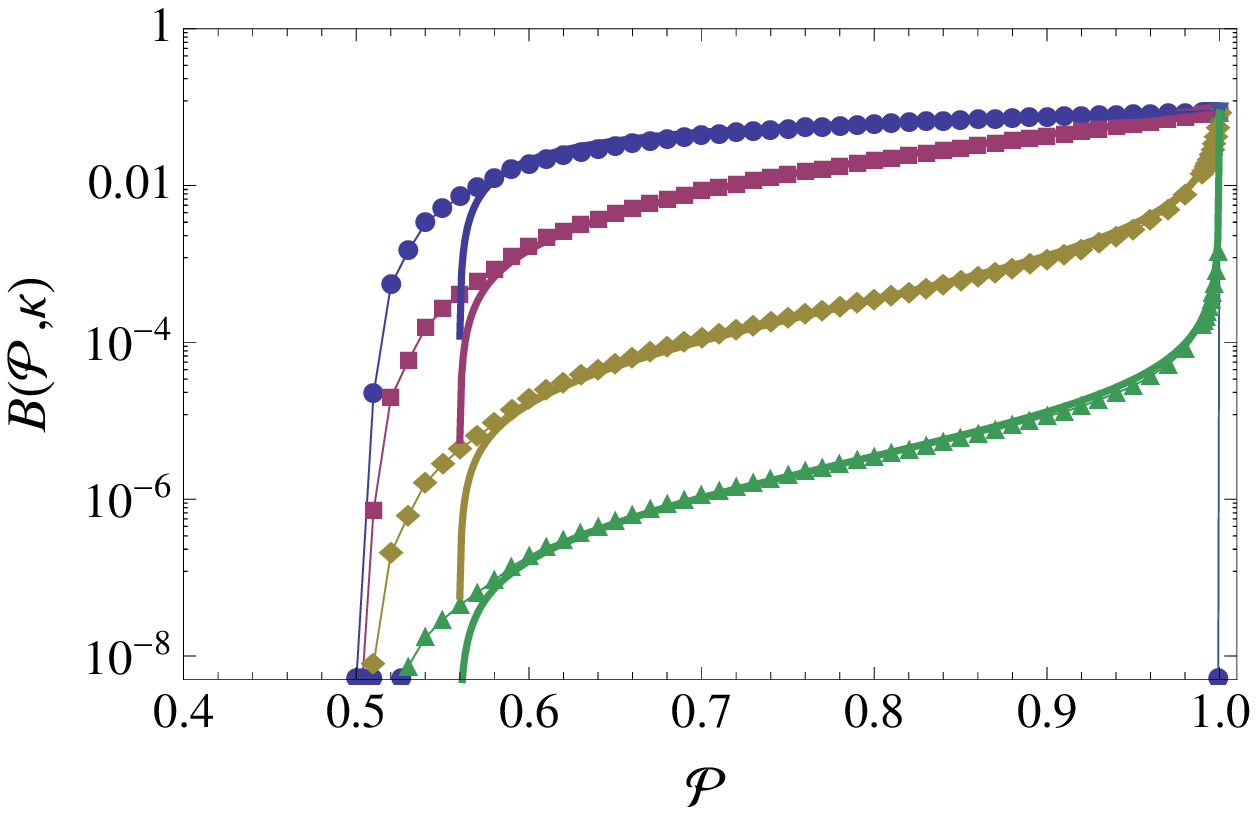}
\end{minipage}
\hspace{0.5cm} \begin{minipage}[b]{0.48\linewidth}
\includegraphics[width=\linewidth]{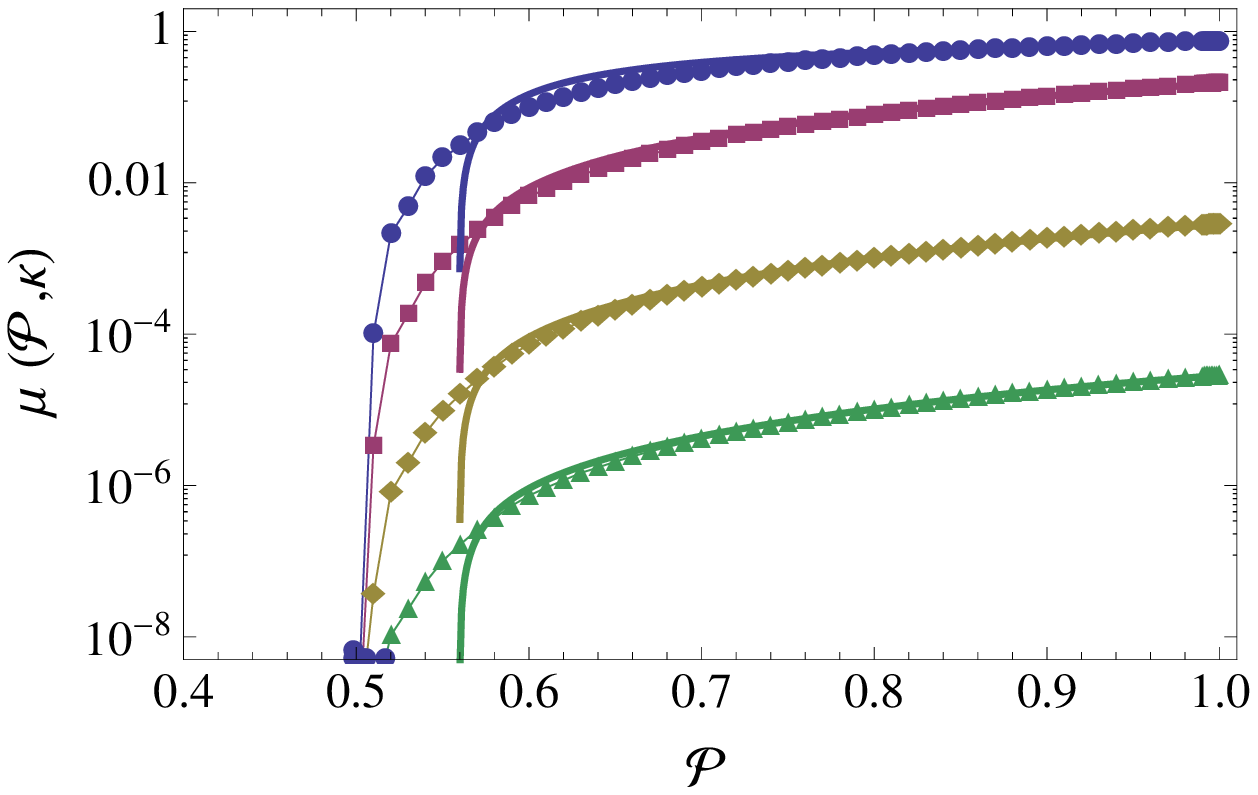}
\end{minipage}
\caption{Bulk (on the left) and shear (on the right) moduli of
the diluted diamond lattice as a function of $\mathcal{P}$, for
$k=1$,
and $\kappa = 1$ (blue circles), $10^{-2}$ (red squares), $10^{-4}$ (yellow diamonds), and $10^{-6}$ (green triangles).}%
\label{gmup-d}%
\end{figure*}

In our EMT, the bulk and shear moduli of the honeycomb and
diamond lattices have the same form as functions of the
effective medium spring and bending constants $k_m$ and
$\kappa_m$:
\begin{eqnarray}
B &=& A_{B}\, k_m = \frac{ 1 }{ 3 } \, (C_{11}+2 C_{12}), \nonumber \\
\mu &=&  A_{\mu}\frac{k_m \kappa_m / \ell_0^2}{\beta k_m +
\gamma\kappa_m / \ell_0^2} 
\nonumber \\
&=& \frac{A_\mu}{\beta} k_m{ \tilde\kappa}_m \left(1+(\gamma/\beta){\tilde\kappa}_m\right)^{-1} = C_{44}
\nonumber \\ &=& 
\left\{
\begin{array}{ll}
\displaystyle\frac{A_\mu}{\beta}\frac{\kappa_m}{\ell_0^2}, & \mathrm{if} \,\, {\tilde\kappa}_m \ll 1,
 \\
\displaystyle\frac{A_\mu}{\gamma} k_m, & \mathrm{if} \,\, {\tilde\kappa}_m \gg 1 ,
\end{array}
\right.
\label{self-consistent}
\end{eqnarray}
where $C_{ij}$ are the standard Voigt elastic constants for a
cubic crystal, $\tilde{\kappa}_m =\kappa_m/( k_m \ell_0^2 )$, $A_B\equiv A_{B,\mathrm{H}}=3/4$, $A_{\mu}\equiv
A_{\mu\mathrm{H}}=27/2$, $\beta\equiv \beta_{\mathrm{H}}=2$, and
$\gamma\equiv \gamma_{\mathrm{H}}=9$ for the honeycomb lattice,
and $A_B\equiv A_{B,\mathrm{D}}=1/12$, $A_\mu\equiv
A_{\mu,\mathrm{D}}=144$, $\beta\equiv \beta_{\mathrm{D}}=27$, and
$\gamma\equiv \gamma_{\mathrm{D}}=192$ for the diamond lattice.
Thus $B$ and $\mu$ are determined as a function of $\kappa$ and
$\mathcal{P}$ once the EMT equations (\ref{eq:k_m}) and
(\ref{eq:kappa_m}) are solved.

Figure \ref{bmup} shows plots of numerical solutions of the EMT
equations (solid lines) and simulations (symbols) for the bulk
(left) and shear (right) moduli of the honeycomb lattice, as a
function of the probability $\mathcal{P}$, for $\kappa =1,\,
10^{-2}, \, 10^{-4}$, and $10^{-6}$ (in blue, red, yellow, and
green respectively). We will keep the same color definitions in
all subsequent plots. Figure \ref{gmup-d} shows similar plots
for the diamond lattice. In both cases, simulations and the EMT
results agree well near $\mathcal{P} = 1$. In the vicinity of
the rigidity threshold $\mathcal{P} = \mathcal{P}_c$, the
simulations display a decay that is different from that found in the EMT. We find $\mathcal{P}_c
\approx 0.5$ in the simulations for the diamond lattice. We can
use Eq.~(\ref{eq:rThorpe}) and,
\begin{eqnarray}
z\mathcal{P} = \left( \sum_{i=0}^z i \, n_i \right) \left/ \left( \sum_{i=0}^z n_i \right. \right),
\end{eqnarray}
along with a combinatorial calculation:
\begin{eqnarray}
n_i =  N
%\binom{z}{i}
{z \choose i}
\mathcal{P}^i (1-\mathcal{P})^{z-i},
\end{eqnarray}
to show that at $\mathcal{P}=\mathcal{P}_c$, $r=r_c \approx
2.5$, which is close to the value $r_c=2.4$ that was obtained
by He and Thorpe~\cite{he85} using a different protocol.

The EMT equations provide analytic expressions for $\kappa_m$
and $k_m$ as a function of $\kappat$ and $\mathcal{P}$ in the
vicinity of $\mathcal{P}=1$ and $\mathcal{P}=\mathcal{P}_c$, where
\begin{eqnarray}
\kappat \equiv \frac{\kappa}{ k \, \ell_0^2},
\end{eqnarray}
is a unitless measure of the bending stiffness. We
begin with $\mathcal{P}$ near one and seek expressions valid at
small $\kappat$ where $\tilde{\kappa}_m$ is
approximately equal to $\kappat$.  Thus we expand $a^*$ and
$b^*$ to linear order in $\tilde{\kappa}_m$. When $\tilde{\kappa}_m = 0$ exactly, $a^*$
is ill defined because $\bm{D}_{s,\bm{q}}$ has a nullspace
[dimension $1$ ($2$) for the honeycomb (diamond) lattice)], and
its inverse does not exist. When $\tilde{\kappa}_m$ is small, the projection
of $\bm{D}_{\bm{q}}$ onto the null space of $\bm{D}_{s,\bm{q}}$
is proportional to $\tilde{\kappa}_m^{-1}$, but the contribution of this
projection to the trace is zero because by definition
$\bm{D}_{s,\bm{q}}$ is zero in its own nullspace. Thus, $a^*
(\tilde{\kappa}_m)$ has a well-defined limit as $\tilde{\kappa}_m \rightarrow 0$ and a
well-defined power series in $\tilde{\kappa}_m$.  To linear order
\begin{equation}
a^*=1-\alpha \, \tilde{\kappa}_m + \mathcal{O}({\tilde{\kappa}_m}^2) ,
\end{equation}
where $\alpha$ is $3.39$ and $4.62$ for the Honeycomb lattice
and diamond lattices, respectively. A similar analysis can be
applied to $b^*(\tilde{\kappa}_m)$, but with $\bm{D}_{b,\bm{q}}$ rather than
$\bm{D}_{s,\bm{q}}$ having a nullspace. It is easier, however,
to use the relation, Eq.~(\ref{abRelation}) between $b^*$ and
$a^*$ to obtain
\begin{equation}
b^*= \frac{2d-z}{z(z-1)} + \frac{\alpha}{z-1} \tilde{\kappa}_m + \mathcal{O}({\tilde{\kappa}_m}^2).
\label{b*}
\end{equation}
For small $\Delta \mathcal{P}=1-\mathcal{P}$ and $\kappat$,
these results imply
\begin{eqnarray}
\frac{k_m}{k} &=& 1 - \frac{\Delta \mathcal{P}}{1-a^*}
\nonumber \\
&=& 1- \frac{\Delta \mathcal{P}}{\alpha \, \tilde{\kappa}_m} + \mathcal{O} (\Delta \mathcal{P} \, \tilde{\kappa}_m)
\nonumber \\
&=& 1- \frac{\Delta \mathcal{P} \, k_m \, {\ell_0}^2}{\alpha \, \kappa} + \mathcal{O} (\Delta \mathcal{P} \, \tilde{\kappa}_m, \Delta \mathcal{P}^2),
\end{eqnarray}
where in the last line we have used Eqs. (\ref{eq:kappa_m}), (\ref{eq:definition_bm}) and (\ref{b*}). Thus,
\begin{eqnarray}
\frac{k_m}{k} &\approx &\left(1+\frac{1}{\alpha \kappat} \Delta \mathcal{P}\right)^{-1}, \\
\end{eqnarray}
and similarly,
\begin{eqnarray}
\frac{\kappa_m}{\kappa} &\approx& 1 - \frac{2 z(z-1)}{z^2 -2 d -z \alpha \kappat } \Delta \mathcal{P} .
\end{eqnarray}

\begin{figure}[th]
\vspace{0.5cm}
\begin{center}
\includegraphics[width=0.9\linewidth]{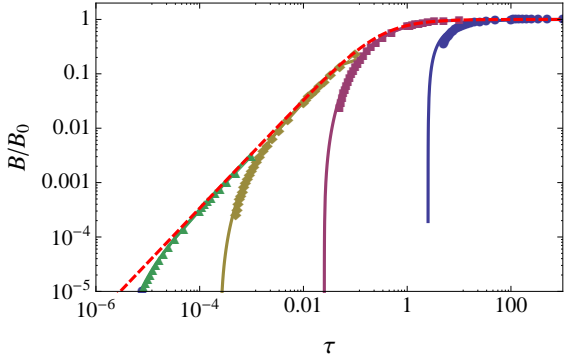}
\end{center}
\caption{Scaling collapse for the bulk modulus of the honeycomb lattice
near $\mathcal{P}=1$. The solid lines are the numerical solutions to the complete
EMT equations, the dashed line the scaling
solution of Eq.~(\ref{scalingFunction}), and the data points are from simulations.}
\label{scalingHoneycombB}
\end{figure}

$\kappa_m$ is a function of $\Delta \mathcal{P}$ and $\kappa$
separately, but $k_m$ is a function of the ratio
\begin{equation}
\tau = \frac{\kappat}{\Delta \mathcal{P}}
\end{equation}
only. We can then define a scaling function $\Xi$ for the bulk
modulus $B$ near $\mathcal{P} = 1$:
\begin{eqnarray}
B = B_0 \, \Xi \left(\tau \right),
\end{eqnarray}
where,
\begin{eqnarray}
\Xi(\tau) = \left(1 + \frac{1}{\alpha \tau} \right)^{-1} 
\label{scalingFunction}
\end{eqnarray}
and $B_0 = A_\mu k$. When $\tau \rightarrow \infty$, [$\Delta \mathcal{P} \ll
\kappa/(k \ell_0^2 )$] , $B$ approaches the undiluted,
$\kappa$-independent limit of $B_0 = A_B k$ characterized by
affine compression of all bonds. When $\tau \rightarrow 0$
[$\Delta \mathcal{P} \gg \kappa/(k \ell_0^2 )$], $B \rightarrow
\kappat/(\alpha \Delta \mathcal{P})$, indicating that bonds in
the diluted lattice are bent in response to isotropic
compression and that response is nonaffine. Figure
\ref{scalingHoneycombB} plots $B/B_0$ as a function of $\tau$
for the honeycomb lattice obtained from full numerical solution
of the EMT equations, the scaling function of
Eq.~(\ref{scalingFunction}), and numerical simulations.  The
three curves agree extremely well near $\mathcal{P}=1$.  In
addition, the simulation and full EMT curves follow each other
closely, and both break away from the scaling curve at
increasing values of $\Delta \mathcal{P}$ as $\kappat$
decreases. The shear modulus $\mu$ cannot be expressed as a
scaling function of $\tau$. It approaches 
\begin{equation}
\mu_0 = (A_\mu /\beta)\, k \, (1+(\gamma/\beta){\tilde \kappa} )^{-1}
\end{equation}
as $\Delta \rightarrow 0$.

\begin{figure*}[th]
\vspace{0.5cm} \begin{minipage}[b]{0.48\linewidth}
\includegraphics[width=\linewidth]{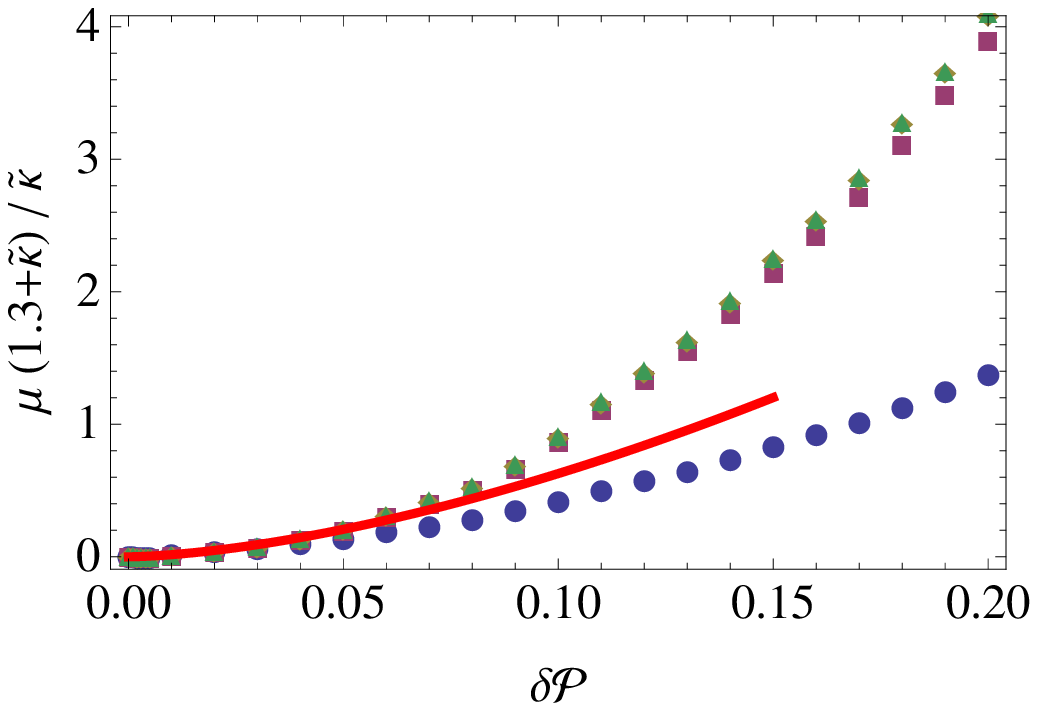}
\end{minipage}
\hspace{0.5cm} \begin{minipage}[b]{0.48\linewidth}
\includegraphics[width=\linewidth]{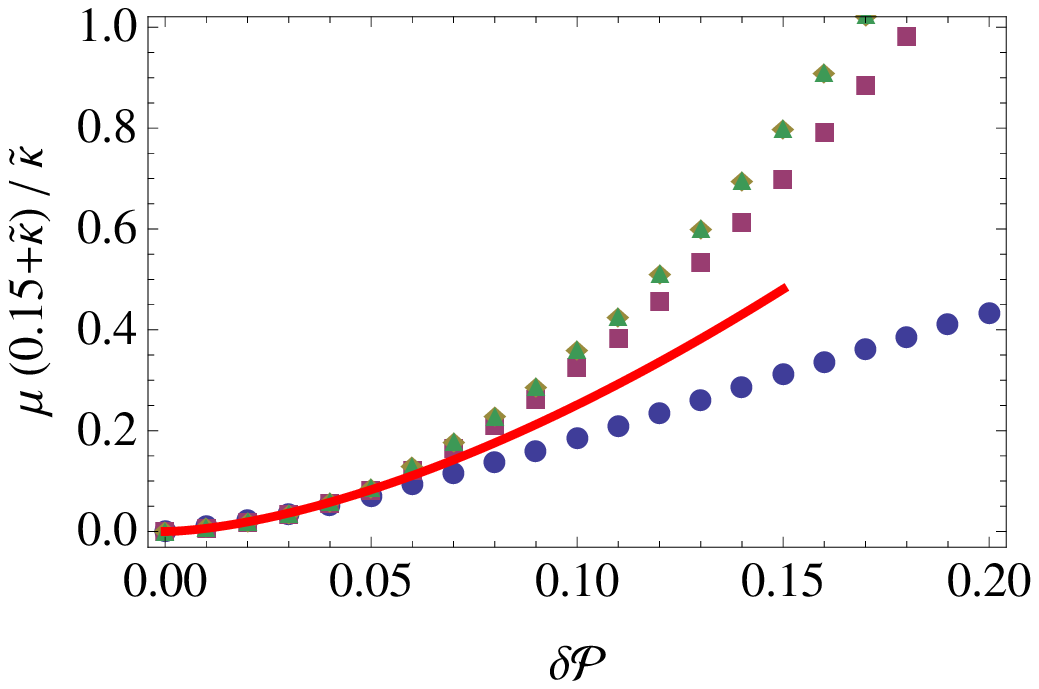}
\end{minipage}
\caption{Scaling collapse for the shear modulus for the honeycomb (left) and diamond (right) lattices
near $\mathcal{P}=\mathcal{P}_c$, for $k=1$, and $\kappa = 1$ (blue circles), $10^{-2}$ (red squares), $10^{-4}$ (yellow diamonds), and $10^{-6}$ (green triangles). The red line is a best fit near $\delta \mathcal{P}=0$. The smaller values at $ {\tilde \kappa} = 1$ are consistent with the EMT expression for $\mu$ (Eq.~(\ref{self-consistent})). The simulations power-law decay is characterized by an approximate exponent of $1.6$ for both lattices.}%
\label{scalingHoneycombMuPc}
\end{figure*}

\begin{figure}[th]
\vspace{0.5cm}
\begin{center}
\includegraphics[width=0.9\linewidth]{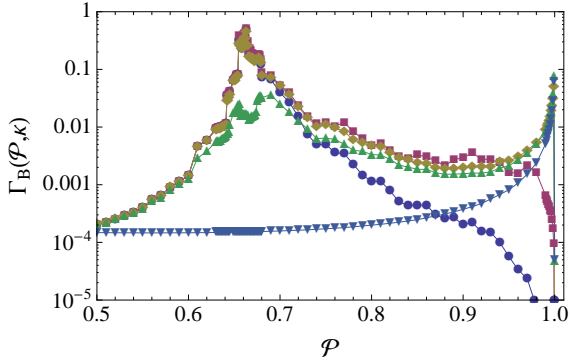}
\end{center}
\caption{Nonaffinity ratio of the bulk modulus $\Gamma_B$
for the honeycomb lattice as a function of the probability $\mathcal{P}$.}%
\label{gamma}
\end{figure}

To obtain analytic expressions for the bulk and shear moduli
near the rigidity threshold at $\mathcal{P} = \mathcal{P}_c$
[given by Eq.~(\ref{abRelation})], we begin with the fact that
$k_m = \kappa_m = 0$ at that point. Thus
\begin{eqnarray}
a^*(\kappa_c)= \mathcal{P}_c, \qquad
b^*(\kappa_c)= \mathcal{P}_c^2 ,
\label{eq:a*b*}
\end{eqnarray}
where $\kappa_c$ ($\approx 0.91$ for the honeycomb lattice and
$\approx 0.30$ for the diamond lattice) is the value of $\tilde{\kappa}_m$
at the rigidity threshold obtained by solving the $a^*(\kappa_c)$
equation for $\kappa_c$ with $\mathcal{P}_c$ given by
Eq.(\ref{abRelation}). We are interested in what happens to
lowest order in $\delta \mathcal{P}= \mathcal{P}-\mathcal{P}_c$
as it increases from zero. To this end, we set $a^* =
\mathcal{P}_c - c(\kappat) \delta \mathcal{P}$ and
$b^*=\mathcal{P}_c^2(\kappat) +c(\kappat) \delta
\mathcal{P}/(z-1)$ with the coefficient $c(\kappat)$ as yet
undetermined.  Then we use Eqs.~(\ref{eq:k_m}) and
(\ref{eq:kappa_m}) to obtain the ratio $\kappa_m/(\kappat \,
k_m \ell_0^2)$ as a function of $\delta \mathcal{P}$. This
ratio must approach $\kappa_c/\kappat$ in the limit $\delta
\mathcal{P} \rightarrow 0$. But because of
Eqs.~(\ref{eq:a*b*}), both the numerator and the denominator of
this ratio are proportional to $\delta \mathcal{P}$ and the
ratio itself depends on $c$ but not $\delta \mathcal{P}$ in the
limit $\delta \mathcal{P} \rightarrow 0$. This limit equation
determines $c(\kappat)$:
\begin{equation}
c(\kappat) = \frac{\kappa_c - 2(z-1) \mathcal{P}_c s_c \kappat}{\kappa_c + \kappat s_c}
\end{equation}
where
\begin{equation}
s_c = \frac{1-\mathcal{P}_c}{(z-1)(1-\mathcal{P}_c^2)}
\end{equation}
The equations for $k_m$ and $\kappa_m$ then become
\begin{eqnarray}
\frac{k_m}{k} \approx \frac{1 - c(\kappat)}{1 - \mathcal{P}_c} \delta \mathcal{P},
\label{kScaling}
\end{eqnarray}
and
\begin{eqnarray}
\frac{\kappa_m }{\kappa} \approx \frac{2(z-1)\mathcal{P}_c + c(\kappat)}{(z-1) (1-\mathcal{P}_c^2)} \delta \mathcal{P} .
\label{kappaScaling}
\end{eqnarray}
Equations (\ref{kScaling}) and (\ref{kappaScaling}) predict
linear behavior of the elastic moduli near $\mathcal{P} =
\mathcal{P}_c$. However, scaling collapse plots of the
simulation results suggest that $B$ and $\mu$ scale with
$\delta \mathcal{P}^{1.6}$, for both honeycomb and diamond lattices, as it is shown in Figure
\ref{scalingHoneycombMuPc} for the shear modulus. The exponent $1.6$,
which is near the value $1.5$ found by He and Thorpe~\cite{he85}, is obtained as the best fit of our data near $\delta \mathcal{P}=0$. However, we note that a log-log plot of this data is very noisy near the rigidity percolation threshold, and that a more precise determination of the exponent requires a careful analysis of finite-size scaling, which is beyond the goal of this manuscript. Finally, the simulation nonaffinity ratio for the bulk modulus $\Gamma_B$,
as a function of $\mathcal {P}$, is shown in Fig. \ref{gamma}.

In our effective medium theory, the ratio of $\mu$ to $B$ is a
function of $\tilde{\kappa}_m$:
\begin{equation}
\frac{\mu}{B} = \frac{A_\mu}{A_B}\left(\frac{\beta}{\tilde{\kappa}_m}+ \gamma \right)^{-1} .
\end{equation}
At the rigidity threshold at which $\tilde{\kappa}_m=\kappa_c$, this ratio
becomes
\begin{equation}
\frac{\mu}{B} =
\left\{    \begin{array}{ll}
    1.608 & \mathrm{Honeycomb} \\
    6.097 & \mathrm{Diamond}
    \end{array} \right.
\end{equation}
The Poisson ratio at threshold of the honeycomb lattice,
\begin{equation}
\sigma_P=\frac{1-(\mu/B)}{1+(\mu/B)} = -0.233 ,
\end{equation}
is thus negative.  The Poisson ratio of the diamond lattice,
which continues to have macroscopic cubic symmetry in the EMT
depends on direction of stresses, and we do not calculate it.

\section{Summary}
To summarize, we have studied phonon and elastic properties of
the honeycomb and diamond lattices with nearest-neighbor
interactions in the harmonic approximation and rotationally
invariant bending interactions. These lattices are
under-coordinated, in the sense that their average coordination
numbers are smaller than the isostatic limit $z_c = 2d$. They
present oscillation modes which reduce to floppy modes over the
entire Brillouin zone when the bending energy coupling constant
vanishes. We implement disorder by assigning a probability
distribution to the existence of each bond in the lattices.
When the number of diluted bonds hits a critical value, there
is a rigidity percolation phase transition at which both bulk and
shear moduli vanish. We employ numerical simulations and an
effective-medium theory to study scaling behavior near
$\mathcal{P}=1$, where the effective-medium theory reproduces
the exact results for the undiluted lattice, and near the
critical probability $\mathcal{P} = \mathcal{P}_c$. The scaling
behavior predicted by the EMT near $\mathcal{P} = 1$ is fairly well
satisfied by the numerical simulations. However, the EMT
predicts a different decay near $\mathcal{P} =
\mathcal{P}_c$, with the elastic moduli proportional to $\delta
\mathcal{P}$, in contrast with the simulation results, where
$B, \mu \sim {\delta \mathcal{P}}^{1.6}$.

\section*{Acknowledgement}
This work was supported by the Brazilian agencies Fapesp and
Capes (DBL), and the NSF under Grants No. DMR-1104707 (TCL) and No.
DMR-1120901 (OS). DBL thanks James Sethna for
useful comments and suggestions.

\appendix

\section{Derivation of the bending energy in terms of
displacements}

This appendix will show that bending potentials can be cast as
a manifestly rotationally invariant combination of displacement
vectors that obey the Keating Rules \cite{keating66}. First
some notation: Let $\rv_\mu$ be the equilibrium reference
position of site $\mu$, and let $\Rv_{\mu}= \rv_\mu + \uv_\mu$
be the position of site $\mu$ after distortion, where $\uv_\mu$
is the displacement vector. Now, consider two bonds, which we
label $a$ and $b$, sharing a common site, which we label as
site $0$. Bond $a$ connects site $1$ to site $0$ and bond $b$
connects site $2$ to site $0$.  Then define
\begin{eqnarray}
\rv_a & = & \rv_1-\rv_0; \qquad \rv_b = \rv_2-\rv_0 ; \\
\Rv_a & = & \Rv_0 - \Rv_1 \equiv \rv_a + \uv_a; \qquad \uv_a = \uv_1 -\uv_0 \nonumber\\
\Rv_b & = & \Rv_2 - \Rv_1 \equiv \rv_b + \uv_b; \qquad \uv_b = \uv_2 -\uv_0 . \nonumber
\end{eqnarray}
$\rv_{a,b}$ is the equilibrium vector and $\Rv_{a,b}$ the
stretched vector for bond $a$ ($b$).

We seek an energy that depends on the angle $\beta_{ab}$
between bonds $a$ and $b$. To define the $\beta_{ab}$, we
assume that it lies between $0$ and $\pi$ so that its sine is
positive (extension to negative $\beta$ is possible but not
relevant to our current interest).  Then
\begin{equation}
\sin \beta_{ab} =\frac{|\Rv_a \times \Rv_b|}{R_a R_b};
\qquad \sin \beta_0 = \frac{\rv_a \times \rv_b}{r_a r_b} ,
\end{equation}
where $R_a = |\Rv_a|$ and $r_a = |\rv_a|$, and we set the
bending energy of the bond-pair $ab$ to
\begin{eqnarray}
E_b & = & \frac{1}{2} \kappat (\beta_{ab} - \beta_0)^2 , \qquad \mathrm{or} \\
& = & \frac{1}{2} \kappat \left(\sin^{-1} \frac{|\Rv_a \times \Rv_b|}{R_a R_b}
- \sin^{-1} \frac{|\rv_a \times \rv_b|}{r_a r_b} \right)^2 .  \nonumber
\label{eq:E_b}
\end{eqnarray}
It is clear that this energy is rotationally invariant.

We can now express these energies in terms of
\begin{eqnarray}
v_a & = &\frac{1}{2}(R_a^2 - r_a^2) = \frac{1}{2}(2 \rv_a \cdot\uv_a +
\uv_a\cdot \uv_a) \rightarrow r_{ai}r_{aj} u_{ij} \nonumber\\
v_b & = &\frac{1}{2}(R_b^2 - r_b^2) = \frac{1}{2}\left(2 \rv_b
\cdot \uv_b+ \uv_b\cdot \uv_b \right) \rightarrow r_{bi}r_{bj} u_{ij} \nonumber\\
v_{ab} & = & \frac{1}{2}\left(\Rv_a \cdot \Rv_b - \rv_a \cdot
\rv_b\right) \nonumber \\
& =& \frac{1}{2}\left(\rv_a \cdot \uv_b + \rv_b \cdot
\uv_a \uv_a\cdot \uv_b\right)
\rightarrow r_{ai} r_{bj} u_{ij} , \label{eq:vab}
\end{eqnarray}
where $i$ and $j$ are Cartesian indices $x,y,z$. The final
forms are the long-wavelength continuum limits with
\begin{equation}
u_{ij} = \frac{1}{2}( \partial_i u_j + \partial_j u_i + \partial_i \uv \cdot
\partial_j \uv )
\end{equation}
the usual rotationally invariant nonlinear Lagrangian strain
tensor and $\uv(\xv)$ the displacement field at space-point
$\xv$.  These limits were obtained using $\rv_0 \equiv \xv$ as
the reference point and $\uv_a \rightarrow r_{ai} \partial_i
\uv(\xv)$ and $\uv_b \rightarrow r_{bi}\partial_i \uv(\xv)$.
$v_a$ and $v_b$ are the forms that normally appear in
central-force models and $v_{ab}$ is the quantity that Keaton
introduces. The continuum limits guarantee that a continuum
elastic energy constructed from the bending energy of
Eq.~(\ref{eq:E_b}) will be a function of $u_{ij}$ only, as it
must be to be rotationally invariant.  It should be noted,
however, that the complete bending energies have terms higher
order in derivatives.

We express the quantities in Eq.~(\ref{eq:E_b}) in terms of
$v_a$, $v_b$ and $v_{ab}$:
\begin{eqnarray}
& |\Rv_a \times \Rv_b|^2  = R_a^2 R_b^2 - (\Rv_a\cdot \Rv_b)^2
= |\rv_a \times \rv_b|^2 + 2 V_{ab} \nonumber\\
& V_{ab} = (r_a^2 v_b + r_b^2 v_a - 2 \rv_a \cdot \rv_b v_{ab})
+ 2(v_a v_b - v_{ab}^2)
\label{eq:axb}
\end{eqnarray}
The form of this expression depends on whether $\rv_a$ and
$\rv_b$ are parallel or not: if they are parallel, $\rv_a
\times \rv_b=0$ and $\beta_0= 0,\pi$.  This limit applies to
models for filamentous lattices
\cite{BroederszMac2014,Head2003,Wilhelm2003}.  The other limit
$\beta_0 \neq 0$ applies to the honeycomb and diamond lattices.
It is important to note that Eq.~(\ref{eq:axb}) and thus the
bending energy $E_b$ depends on both the Keating part $v_{ab}$
and the ``central-force" parts $v_a$ and $v_b$, but by
construction it depends only on the the rotation angle between
bonds and not on any stretch in the bonds.  The Keating energy
depends on stretching as well as bending and contributes to
both the bulk and shear moduli. The bulk modulus does not
depend on a pure bending energy of the type we discuss. There
is no change in the angle between bonds under uniform
compression.  With the above definitions,
\begin{equation}
\frac{|\Rv_a \times \Rv_b|}{R_a R_b \sin\beta_0} =
\left(\frac{1+2 V_{ab}/|\rv_a \times \rv_b|^2}{(1+ 2 v_a/r_a^2)(1+2 v_b /r_b^2)}
\right)^{1/2} \\
\end{equation}
and
\begin{equation}
\beta - \beta_0\approx \tan\beta_0 \left(\frac{V_{ab}}{|\rv_a \times \rv_b|^2}
-\frac{v_a}{r_a^2} - \frac{v_b}{r_b^2} \right)
\end{equation}
in the small-displacement limit.

Equations Eq.~(\ref{eq:E_b}) and (\ref{eq:axb}) provide a
complete expression for bending energies in terms of nonlinear
functions of the nonlinear ``discrete" strain functions $v_a$,
$v_b$, and $v_{ab}$.  We are often interested in the harmonic
limit of these functions.  We begin with the case $\beta_0
>0$, and we expand to lowest order in $\uv_a$ and $\uv_b$
\begin{equation}
v_a \rightarrow \rv_a \cdot \uv_a, \qquad
v_{ab}\rightarrow \frac{1}{2}(\rv_a\cdot \uv_b + \rv_b \cdot \uv_a ) ,
\label{eq:vavbe}
\end{equation}
and
\begin{eqnarray}
V_{ab} &\rightarrow & r_a^2 \rv_b \cdot \uv_b + r_b^2 \rv_a \cdot \uv_a
- \rv_a \cdot \rv_b (\rv_a \cdot\uv_b + \rv_b \cdot \uv_a) \nonumber\\
& = & r_b^2 \rv_a \cdot \uv_a^{\perp b} + r_a^2 \rv_b \cdot \uv_b^{\perp a} ,
\label{eq:Vabe}
\end{eqnarray}
where $\uv_a^{\perp b}$ is the projection of $\uv_a$ onto the
space perpendicular to $\rv_b$, i.e. $u_{ai}^{\perp b} =
P^b_{ij} u_{aj}$, where $P^b_{ij}= \delta_{ij} -
\hat{r}_{bi}\hat{r}_{bj}$, with $\hat{\rv_b} = \rv_b/r_b$, is
the projection operator onto the plane perpendicular to
$\rv_b$. Then with the aid Eqs.~(\ref{eq:vavbe}) and
(\ref{eq:Vabe}) and the relation
\begin{eqnarray}
&& \left(\frac{1}{\sin^2\beta_0} - 1 \right) \hat{r}_{ai}
P^b_{ij} u_{aj}- \hat{r}_{aj} u_{aj}
\nonumber \\ && \qquad
=\frac{\cos\beta_0}{\sin^2 \beta_0} \hat{r}_{bi}P^a_{ij} u_{aj} ,
\end{eqnarray}
the energy $E_b$ to harmonic order is
\begin{equation}
E_b^{\mathrm{har}} = \frac{\kappat}{2 \sin^2 \beta_0}
\left(\hat{\rv}_b \cdot \frac{\uv_a^{\perp a}}{r_a} + \hat{\rv}_a \cdot \frac{\uv_b^{\perp b}}{r_b}
\right)^2
\end{equation}
Note that this energy depends only on displacements
perpendicular to equilibrium bond directions and thus it does
not induce any bond compression. Setting $\kappat = \kappa \sin
\beta_0)$, $a=lk$, $b=lm$, and  $r_a = r_b = l_0$ leads to
Eq.~(\ref{ebend-b}). If only the $v_{ab}$ part were kept,
$E_b^{\mathrm{har}}$ reduces to the form used in reference
\cite{he85}.

When $\rv_a= r_a \ev$ and $\rv_b= -r_b \ev$ are anti-parallel,
$\beta_0=\pi$,  $\sin^2 \beta= 2 V_{ab}/R_a^2 R_b^2$, and the
linear part of $V_{ab}$ vanishes:
\begin{eqnarray}
V_{ab}^{(1)}  &=& r_b^2 r_a \ev\cdot \uv_a -r_a^2 r_b \ev\cdot \uv_b  
\nonumber \\ && \quad 
+
r_a r_b (r_a \ev\cdot \uv_b - r_b \ev \cdot \uv_a)
= 0.
\end{eqnarray}
The quadratic part of $V_{ab}$ is
\begin{equation}
V_{ab}^{(2)} = \frac{1}{2} (r_a \uv_b^{\perp} + r_b \uv_a^{\perp})^2
\label{eq:Vab-nl}
\end{equation}
When $\uv_{a i}^{\perp} = (\delta_{ij}-e_i e_j)u_{a,j}$, and
$E_b$ becomes $\kappat V_{ab}^{(2)}/(r_a^2 r_b^2)$ to harmonic
order. If we had chosen $\rv_b = \rv_0-\rv_2$ rather than
$\rv_b = \rv_2 - \rv_0$, then $\beta_0 = 0$, and there would be
a minus sign in Eq.~(\ref{eq:Vab-nl}).

%\bibliography{/Users/daniloliarte/Documents/Latex/biblio}% Produces the bibliography via BibTeX.
\section*{References}

%\bibliography{biblio}
%\bibliographystyle{unsrt}

\end{document}